# High-precision control of static magnetic field magnitude, orientation, and gradient using optically pumped vapour cell magnetometry

S. J. Ingleby[1,2], P. F. Griffin[1], A. S. Arnold[1], M. Chouliara[1], E. Riis[1]

[1] Department of Physics, SUPA, Strathclyde University, 107 Rottenrow East, Glasgow, UK
[2] stuart.ingleby@strath.ac.uk

**Abstract**

An integrated system of hardware and software allowing precise definition of arbitrarily oriented magnetic fields up to |B| = 1 µT within a five-layer mumetal shield is described. The system is calibrated with reference to magnetic resonance observed between Zeeman states of the $6S_{1/2}$ F = 4 $^{133}$Cs ground state. Magnetic field definition over the full 4π solid angle is demonstrated, with one-sigma tolerances in magnitude, orientation and gradient of δ|B| = 0.94 nT, δθ = 5.9 mrad and $\delta|\nabla B|$ = 13.0 pT/mm, respectively. This field control is used to empirically map $M_x$ magnetometer signal amplitude as a function of the static field ($B_0$) orientation.

## I. Introduction

Alkali vapour cell magnetometry offers extremely high precision B-field measurements [1,2], but the response and resolution of finite-field vapour cell magnetometers is dependent on field orientation [3]. The study of magnetometer response to arbitrarily oriented fields could be of significant benefit in the design of practical sensors for surveying [4], geophysical [5] and medical applications, such as magnetoencephalography (MEG) [6] and magnetocardiography (MCG) [7]. Double resonance magnetometry offers the advantage of absolute measurement of |B|, without necessitating sensor calibration [8,9], and has important applications in field control for fundamental physics, such as neutron electric dipole moment searches [10].

We draw a broad distinction between optically pumped magnetometry schemes in which magnetic field modulation is *(1)* small and applied at a frequency resonant with atomic Larmor precession due to the measured field $B_0$ (known as *double-resonance* magnetometry) [11], and *(2)* greater than $B_0$, and applied at low frequency, allowing detection of zero-field resonances (*zero-field* magnetometry) [12]. The highest sensitivities have been achieved using zero-field magnetometry in shielded environments, where narrow-linewidth resonances may be observed in the spin-exchange relaxation free (SERF) regime for B-field magnitudes below ~10 nT [13]. In contrast, double-resonance magnetometry can be used to measure geophysical fields without compensation [14], and, using modulation of light frequency [15], polarisation [16] or amplitude [17], may be implemented without the use of applied alternating fields. In cases where the modulation is applied using a field perturbation $B_{RF}$, a further distinction can be drawn between magnetometers measuring resonant precession of atomic spin orientation (in which the pump light creates vector polarisation in the sample) and alignment (in which the pump light creates tensor polarisation in the sample).

In this work we study a magnetometry scheme which might usefully be applied in a portable, compact, unshielded sensor. A single circularly-polarised monochromatic light source is used, and the system is weakly modulated by a small field perturbation, allowing measurement of the ambient field $B_0$. Progress in the application of such sensors for unshielded measurements could be furthered by empirical study of $B_0$ orientation and dead-zone effects. In order to investigate these effects, a test system was constructed in which a homogenous B-field of precisely defined magnitude and orientation was established within a shielded environment.

In order to achieve precise control of B-field magnitude and orientation, a software-controlled field optimisation routine was developed, taking advantage of double-resonance magnetometry to measure absolute field magnitude and implement a self-calibrating system. In addition to nT-level compensation, which has been reported using Hanle effect zero-field techniques [18], [19], the method presented here demonstrates minimisation and control of the first-order magnetic field gradient $\nabla B$.

First we describe the electronic and optical apparatus, and the double-resonance measurement used to determine the field magnitude and gradient. We then describe the iterative routines developed to accurately determine the null field, coil response and field gradient compensation, and evaluate the isotropy of the field generated by the calibrated system. Finally we demonstrate the measurement of anisotropy in orientation-based double-resonance ($M_x$) magnetometry using the calibrated system.

## II. Hardware

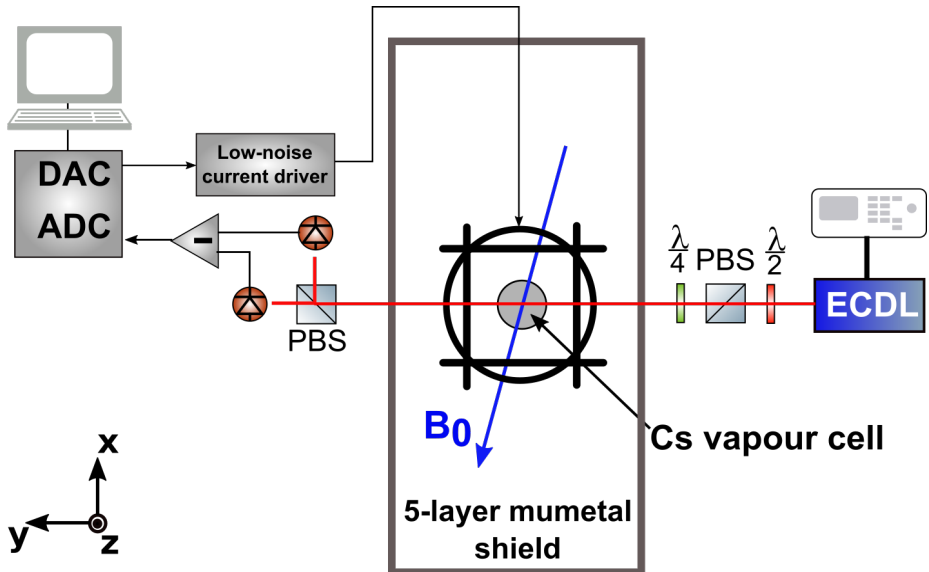

**Figure 1:** Schematic showing experimental system, including extended-cavity diode laser (ECDL), polarisation optics, digital-to-analogue (DAC) and analogue-to-digital (ADC) converters.

Figure 1 shows the test system. Approximately $10^{-6}$ torr ($\sim 3.10^{10}$ cm$^{-3}$) $^{133}$Cs vapour is contained within a spherical paraffin-coated cell of diameter 28 mm. This cell is maintained at room temperature (20.7 °c) throughout the measurements. In these conditions the optical thickness of the unpolarised vapour is $\sim 0.6$. No additional buffer gasses are used, but Cs polarisation lifetime is extended to > 100 ms through suppression of spin-relaxing Cs-wall collisions by a paraffin layer on the inside of the cell wall [20]. An extended-cavity diode laser (New Focus Vortex 6800), on resonance with the Cs 895 nm D1 F = 4 to F = 3 transition, is used to create spin orientation in the Cs sample through optical pumping with circularly polarised light propagating along *y*.

The Cs cell is housed within a three-axis Helmholtz coil assembly, comprising six independently-controlled coils of radius and separation 49 mm, 35 mm and 42 mm on the *x*, *y* and *z* axes, respectively. Additional Helmholtz coils are wound on the *y* and *z* axes in order to apply a small oscillating modulation field $B_{RF}$. The coil assembly is a neat fit inside the innermost of five mumetal shields, which has an inner diameter of 100 mm and aspect ratio of 8:1. Access to the coils inside the shields is provided via 25 mm axial ports and 10 mm transverse ports.

Each shield is wound with a ten-turn degaussing coil, capable of saturating the shield with an applied current of 0.4 A. Software-controlled current supplies are used to sequentially degauss the shields (inner to outer) using a 5 Hz alternating current whose amplitude exponentially decays from 0.5 A to 1 mA over a 12 min period. This sequence is repeated 5 times to ensure that the DC shielding factor is maximised and residual flux trapped in the innermost shield is minimised [21].

Current supply for the Helmholtz coil assembly is provided by an eight-channel circuit built around the high-current OPA551 amplifier, which provides currents up to 200 mA with 3.5 $fA.Hz^{-1/2}$ noise floor at 1 kHz bandwidth. The output current of this amplifier is controlled by an input reference voltage, provided by a 13-bit DAC (NI PCI-6723) (see Figure 2). A four-pole low-pass Butterworth filter with -3 dB roll-off at 1 kHz is employed to remove DAC noise from the input voltage. The resolution of the output current (single bit change in the reference voltage) is 2.4 µA, corresponding to a field step of 0.15 nT or gradient step of 4 pT/mm on the z-axis. The maximum reference voltage (±10 V) corresponds to an applied field of ±1.2 µT.

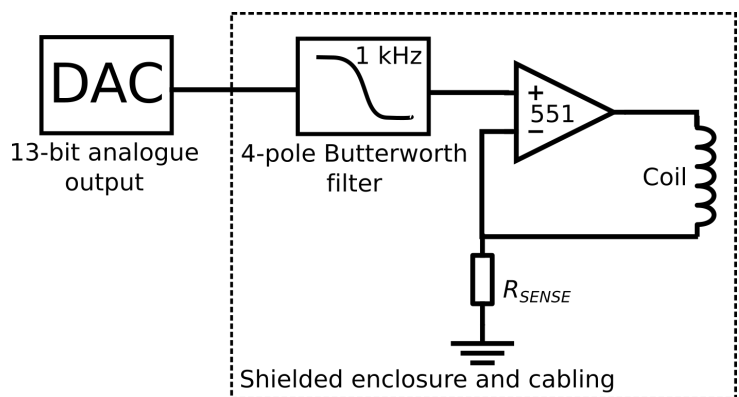

**Figure 2: Schematic of current driver circuit used for software field control. Six independent amplifier circuits are used to set the field magnitude, orientation and gradient inside the shield.**

### III. Magnetic resonance measurement

Optical pumping with a circularly polarised beam of power 10 µW and diameter 2 mm creates a net magnetisation **M** in the Cs sample through orientation of atomic spins (see Figure 3). The evolution of **M** in a time-varying field B(t) can be determined classically using

$$\dot{M} = \gamma M \times B(t) - \Gamma(M - M_{EQ}) \qquad (1)$$

where $M_{EQ}$ is the equilibrium magnetisation and $\Gamma$ is the total spin relaxation rate, in general a tensor including terms for spin dephasing due to B-field gradients, spin-exchange and spin-destruction interactions, and continuous-wave optical pumping. In this case we consider an isotropic single-valued spin relaxation rate in the presence of weak optical pumping, approximating $\Gamma \approx \Gamma I$.

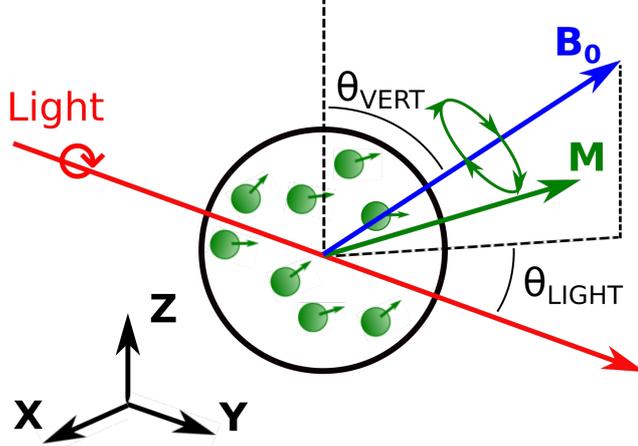

**Figure 3:** Schematic of magnetometry scheme under study, showing classical evolution of vector magnetisation M and the coordinate system used to define $B_0$ orientation.

We consider B(t) to be comprised of a static field $B_0$ and a small alternating field perturbation $B_{RF}$ of frequency $\omega_{RF}$. Under resonant perturbation, in which $\omega_{RF}$ is equal to the Larmor frequency $\omega_L$, M undergoes precession about the axis defined by the static field $B_0$. The Larmor frequency is given by $\omega_L = -\gamma |B_0|$, where $\gamma \approx$ 3.50 Hz/nT for the $6S_{1/2}$ F = 4 ground state of the Cs atom. Contributions due to the nonlinear Zeeman effect are negligible in the low field regime ($2\omega_L^2/\delta\omega_{hfs}$ = 50 µHz [22]), meaning that heading errors will not significantly distort the isotropy of measured B-field magnitudes.

As the optical absorption and dispersion properties of the sample vary with **M**, spin precession can be detected by measurement of polarisation rotation in transmitted light. A balanced polarimeter with noise floor 0.18 µrad.Hz$^{-1/2}$ at 1 kHz is used to detect the response in **M** to $B_{RF}(\omega_{RF})$. For small $B_{RF}$, demodulation of the rotation signal at $\omega_{RF}$ yields the in-phase (X) and quadrature (Y) components of the rotation signal, which are described by dispersive and absorptive Lorentzian lineshapes, centred on $\omega_L$:

$$X = Ax\left[\frac{1}{1+x^2}\right] \qquad (2)$$

$$Y = A\left[\frac{1}{1+x^2}\right] \qquad (3)$$

where A is the on-resonance signal amplitude. If $B_{RF} \ll \Gamma/\gamma$, the Lorentzian half-width-half-maximum (HWHM) is approximately equal to $2\pi\Gamma$ and $x \simeq \frac{\omega_L - \omega_{RF}}{2\pi\Gamma}$ [23]. Alternatively, one may also consider the signal magnitude (R) and phase ($\phi$), given by

$$R = \sqrt{X^2 + Y^2} = A\frac{1}{\sqrt{1+x^2}} \qquad (4)$$

$$\tan(\varphi + \varphi_0) = X/Y = x, \qquad (5)$$

where $\phi_0$ is the on-resonance phase of the signal.

The modulation field $B_{RF}$ and polarisation response are generated and measured synchronously using a 1.25 MHz 16-bit DAQ card (NI PCIe-6353). $X_M(\omega_{RF})$ and $Y_M(\omega_{RF})$ are measured by demodulating the polarimeter response to a modulation field applied for $t_{SAMPLE}$ preceded by $t_{PRETRIGGER}$. $t_{SAMPLE}$ is approximately equal for each $\omega_{RF}$ but, in order to minimise errors in demodulation, is allowed to vary by small increments such that it includes an integer number of modulation periods. Each $t_{SAMPLE}$ commences on an RF modulation phase of zero and demodulation is carried out in software at a phase chosen by the user. By sweeping $\omega_{RF}$ through a fixed range centred on the expected resonance frequency $\omega_0 = -\gamma |B_0|$, and demodulating at a phase equal to $-\phi_0$ (Eq. (5)), resonance curves such as Figure 4 can be recorded. Figure 4 shows magnetic resonance response for an optimised orientation of $B_0$ ($\theta_{LIGHT} = \theta_{VERT} = \pi/4$ rad).

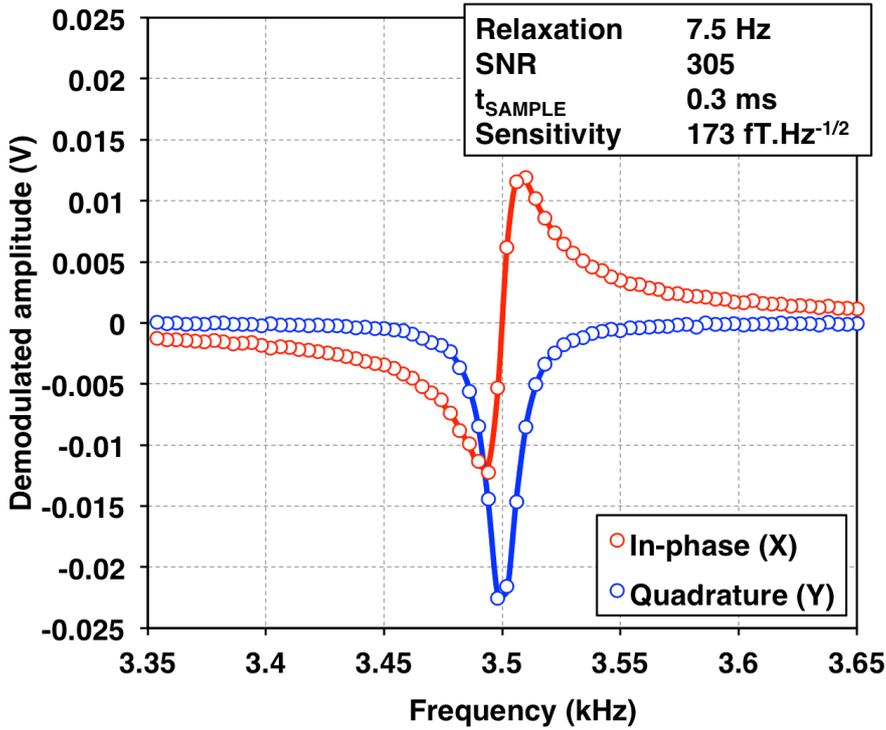

**Figure 4: Magnetic resonance observed in Cs spin precession response to 1 nT amplitude modulation. $B_0$ = 1 µT. Sensitivity is estimated by square root of average noise equivalent power over sensor bandwidth. The resonance is centred on the Larmor frequency $\omega_L/2\pi$, the half-width half-maximum is approximately equal to the relaxation rate $\Gamma$, and the amplitude A varies proportionately with the RF modulation depth and laser power.**

Least-squares fits of Equations (2) – (5) to measured resonance data are used to estimate values and uncertainties for A, $\Gamma$, $\omega_L$ and $\phi_0$. A sequential fitting technique is used; first A, $\Gamma$ and $\omega_L$ are estimated by fitting Eq. (4) to

$$R_M(\omega_{RF}) \equiv \sqrt{X_M(\omega_{RF})^2 + Y_M(\omega_{RF})^2}. \qquad (6)$$

$\phi_0$ is then found by fitting (5) to

$$\varphi_M(\omega_{RF}) \equiv \tan^{-1}(X_M(\omega_{RF})/Y_M(\omega_{RF})). \qquad (7)$$

The phase of the data is then rotated by $-\phi_0$ in a manner analogous to [24] to allow a fit to $X_M(\omega_{RF})$ using (2).

The fitted parameters can be used to estimate magnetic field resolution and sensitivity. On resonance, response to small changes in $B_0$ is given by

$$\frac{dX}{dB} \cong \frac{\gamma A}{\Gamma} \qquad (8)$$

for small modulation amplitudes. Magnetometric resolution $\delta B_{RMS}$ is estimated using the in-phase signal-to-noise ratio *SNR*. *SNR* is found by calculating the ratio of *A* (estimated by fitting (2) to $X_M$) to the root-mean-square value of measured residuals to that fit.

$$\delta B_{RMS} = \frac{dB}{dX} \delta X_{RMS} = \frac{\Gamma}{\gamma \, SNR} \qquad (9)$$

A conservative estimate of magnetic sensitivity is obtained by assuming that resolution is limited by noise equally distributed across the sensor bandwidth.

## IV. Coil calibration and null field compensation

In order to achieve precise control of the magnitude and orientation of the field inside the shield, the field response of the Helmholtz coils to software control was measured and calibrated by observing the magnetic resonance frequency of the Cs atomic sample. The presence of high-permeability shielding material in close proximity to the Helmholtz coil assembly makes accurate calibration of the generated field necessary. A design response was calculated for the coils using Helmholtz' formula, and this value was used in software to establish an expected applied field $B_0$.

In order to measure the null-field compensation *ε* required to cancel any residual background field remaining after degaussing, and the coil calibration factor *a*, defined as the ratio ($B_{measured}$ / $B_{calculated}$), $B_0$ was varied by $\delta B$ on each axis in turn. By magnetic resonance measurement, the measured field magnitude $B_M$, defined $B_M \equiv \omega_L/\gamma$, was found, allowing *a* and *ε* to be found by fitting

$$B_M(\delta B_i) = a_i(\delta B_i - \varepsilon_i) + B_0 \qquad (10)$$

to the measured data (Figure 5). This procedure was carried out using **$B_0$** aligned with the coil under calibration and $B_{RF}$ applied on an axis perpendicular to **$B_0$**. Although the magnetic resonance signal amplitude is suppressed for $B_0$ orientations parallel or perpendicular to the light propagation direction, in practice this condition is never exactly met and there is sufficient signal-to-noise ratio to resolve the magnetic resonance response during the optimisation process, even if the orientation of $B_0$ is close to the light propagation axis.

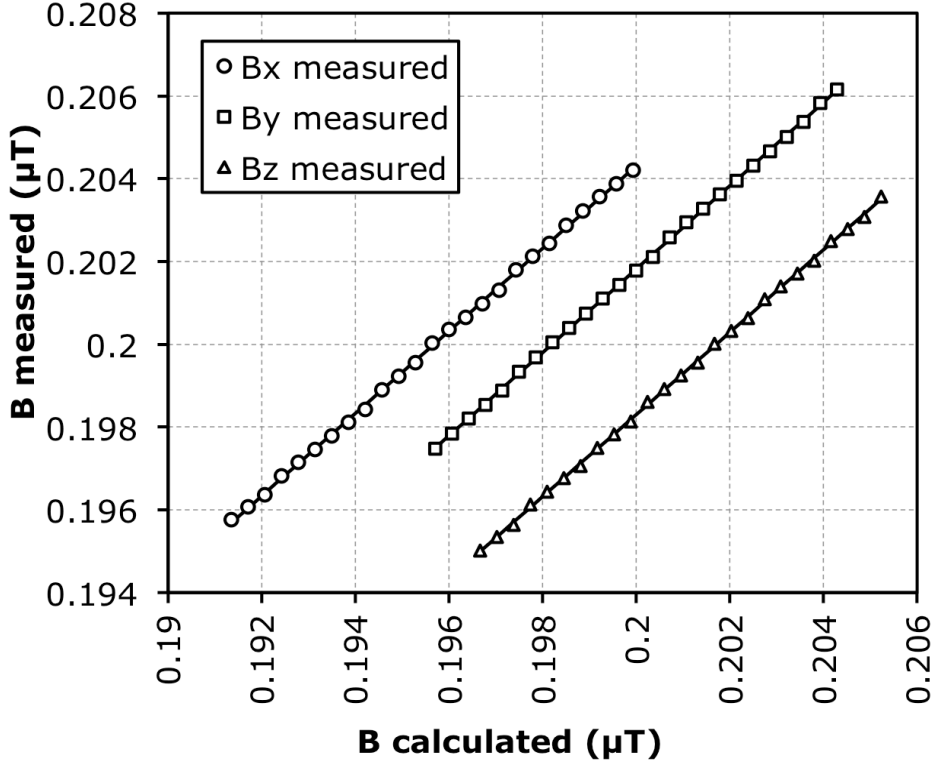

**Figure 5:** Calibration data for coil response and null-field cancellation. The field magnitude is measured as a function of the field $B_{Calculated}$ applied on each axis in turn, while the field components on perpendicular axes are minimised. Results of final iteration for each axis are shown.

| Coil pair | Null field $\varepsilon$ (nT) | Calibration factor $a$ ($B_{measured}$ / $B_{calculated}$) |
|---|---|---|
| X | -4.332(10) | 1.1244(39) |
| Y | -1.799(8) | 1.1800(27) |
| Z | 1.707(10) | 1.3281(35) |

**Table 1:** Fitted parameters and uncertainties for final iteration of coil calibration. The quoted calibration factors are cumulative.

To establish accurate coil calibrations and null-field alignment, several iterations of this procedure are carried out on each coil axis. Figure 5 shows the measured data and Table 1 shows fitted parameters for the final iteration carried out. For each coil, the presence of a high-permeability return yoke resulted in the calibrated coil response being significantly enhanced over the design estimate ($|a| > 1$). The degree of enhancement is dependent on the geometry of the coil relative to the shield axis (the shield axis is aligned along $x$) and proximity of the mumetal to the coil (coil $y$ is the smallest diameter).

The maximum orientation uncertainty for an applied field $B_0$ can be estimated by assuming that the total field uncertainty $\delta B$ is orthogonal to $B_0$ (in the worst case), and is equal to the uncertainties in coil response $\delta a$ and null-field $\delta\varepsilon$ combined. The maximum orientation uncertainty is therefore estimated by

$$\delta\theta \leq \frac{|\delta B|}{|B_0|} \approx \frac{|\delta\varepsilon|}{|B_0|} + |\delta a| \qquad (11)$$

to first order and, for $B_0 = 200$ nT, has an amplitude of 5.9 mrad.

**V. Field gradient minimisation**

The low vapour pressure and absence of buffer gas mean that polarised Cs atoms disperse rapidly and coherently sample the B-field over the entire cell volume. Inhomogeneity in the field across the cell increases gradient contributions to the spin relaxation rate $\Gamma$. This contribution has been previously demonstrated to exhibit a quadratic dependence on the magnitude of the field gradient in the cell volume (see [25]).

By applying a field $B_0$ = 200 nT along $x$ and small field gradient corrections $\nabla B_i^{APP}$ on each axis in turn, field inhomogeneity due to residual flux trapped in the shielding can be compensated and spin relaxation minimised. $\Gamma_M$ is estimated by fitting Eq. 2 to $X_M$. A quadratic model

$$\Gamma_M = \Gamma_0 + b_i(\nabla B_i^{RESID} - \nabla B_i^{APP})^2 \qquad (12)$$

is fitted to $\Gamma_M$ and $\nabla B^{RESID}$ determined. Each gradient optimisation is preceded by redetermination of the null field $\varepsilon$ and coil calibration $a$ by the method described above. Figure 6 shows measurements of $\Gamma_M(\nabla B_i^{APP})$ and fits for the final iteration of this procedure. The magnitude of the one-sigma statistical uncertainty in $|\nabla B^{RESID}|$ after the final iteration is 13.0 pT/mm.

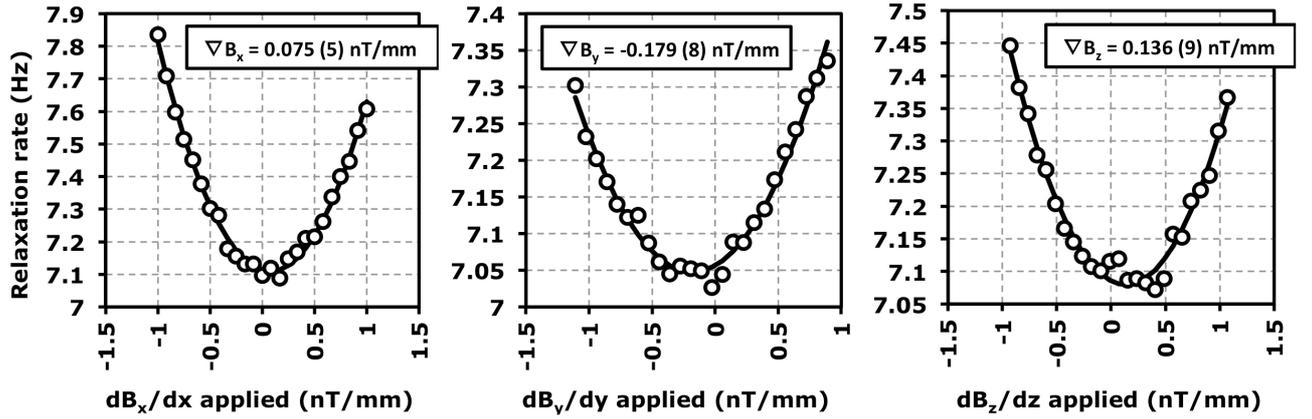

**Figure 6: Measurement of spin relaxation rate with field gradient applied on each coil axis in turn. Orthogonal field gradient components are minimised for each measurement. Applied gradient components for optimum cancellation of background inhomogeneity are shown.**

We note that $\Gamma$ varies more under changes in the field gradient applied along $x$ (the direction of $B_0$) than under gradients applied along $y$ or $z$, indicating good agreement with the results reported in [25]. The discrepancy in $\Gamma_0$ observed between gradients applied on each axis is a consequence of the sequential iterative optimisation routine, which optimises $\nabla B_x^{APP}$, then $\nabla B_z^{APP}$, followed by $\nabla B_y^{APP}$.

## VI. Uniformity of response

After calibrating the coils and compensating for residual field and field gradients, the uniformity of A and $B_M$ can be measured for $B_0$ swept over the full $4\pi$ solid angle with a granularity of $\delta\theta = \pi/36$ rad (Figure 7). $B_{RF}$ is applied on the z-axis. The distribution of $B_M$ is approximately symmetrical around $\langle B_M \rangle$ = 200.12 nT, with a standard deviation of 0.94 nT (Figure 8).

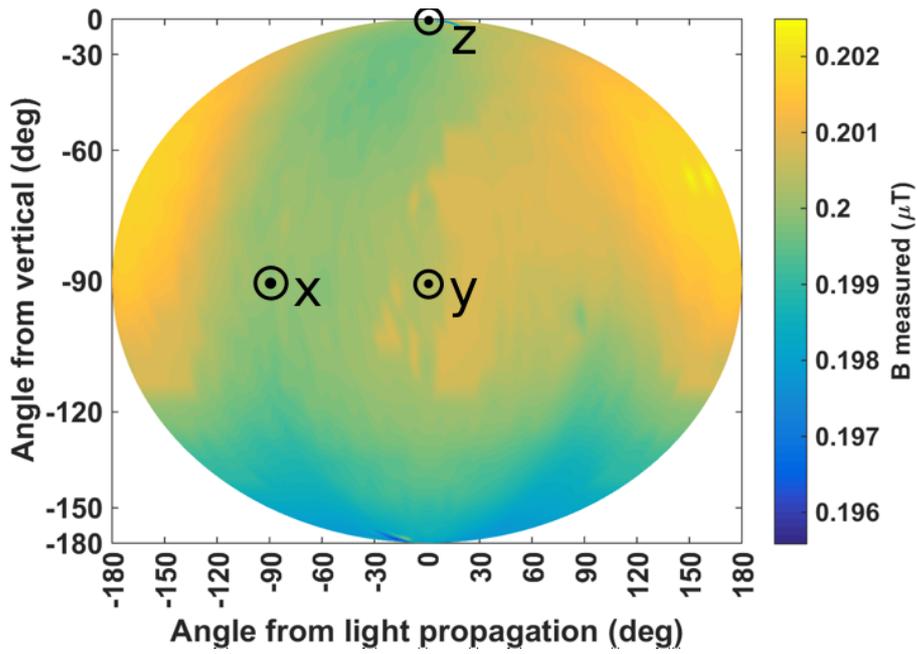

Figure 7: $B_M$ over the full $4\pi$ solid angle. 1646 orientations of $B_0$ are distributed evenly over $4\pi$ solid angle. The average measured field is 200.12 nT with a standard deviation of 0.94 nT. Axis directions are indicated with arrowhead symbols. The coordinate system is shown explicitly in Figure 3.

The data shown in Figure 7 exhibits a slight anisotropy; the deviation from $B_M = 200$ nT, although small, is not randomly distributed with $B_0$ orientation. This is also seen in the non-normal distribution of $B_M$ values (Figure 8).

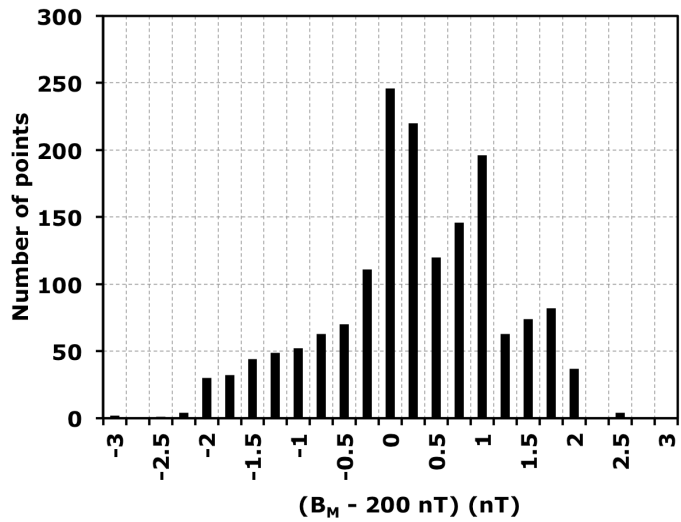

Figure 8: Distribution by $B_M$ value of data points plotted in Figure 7.

The observed anisotropy is consistent with the observation that the limiting factor in $B_0$ definition is the systematic effect of uncertainty in calibration factor *a* (see Table 1) rather than field resolution. $B_0$ is measured with an integration time of 50 ms at each field orientation, and with sensitivity in the sub-pT.Hz$^{-1/2}$ range (see Figure 4), typical $B_M$ resolution can be expected in the pT.Hz$^{-1/2}$ range. The average uncertainty in $B_M$ for the data shown in Figure 7 is $\langle \delta B_M \rangle = 43.1$ pT, whereas an estimate for the uncertainty in B as a result of uncertainty in the coil calibration, $\delta B \approx B_0 |\delta a|$, using the estimates for $\delta a$ given in Table 1, yields $\delta B \approx 1.18$ nT, consistent with the asymmetric spread of $B_M$ in Figure 8.

## VII. Resonance amplitude anisotropy

To demonstrate the type of measurement possible with this precise software control of $B_0$, the resonance amplitude A can be plotted as a function of $B_0$ orientation. Figure 9 shows these data, revealing the highly anisotropic magnetometer response. The observed symmetrical cone-shaped distribution of signal amplitude and dead zones is in approximate agreement with the results of numerical Runge-Kutta integration of Eq. 1 (Figure 10). However, the measured data includes generation and evolution of atomic alignment (as well as higher-order multipole moments), and the anisotropic dephasing effect due to optical pumping, neither of which is represented in Eq. 1. These effects break the rotational symmetry of the magnetometer response about the *y* axis.

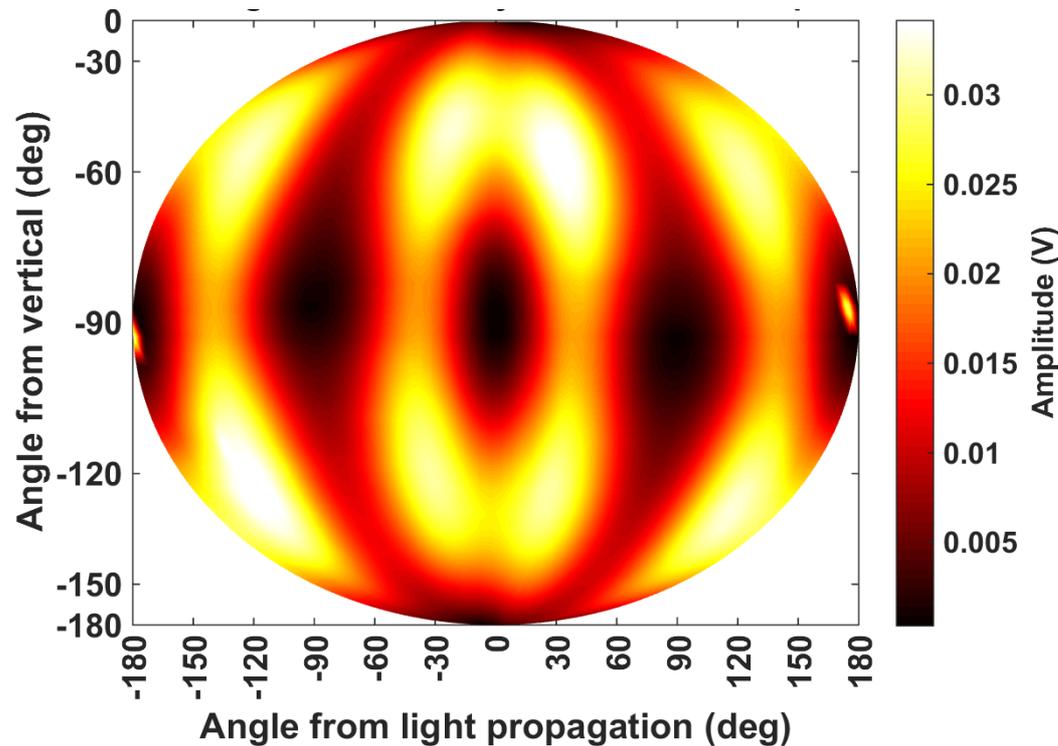

**Figure 9: Anisotropy in amplitude of the magnetic resonance signal with varying $B_0$ orientation.**

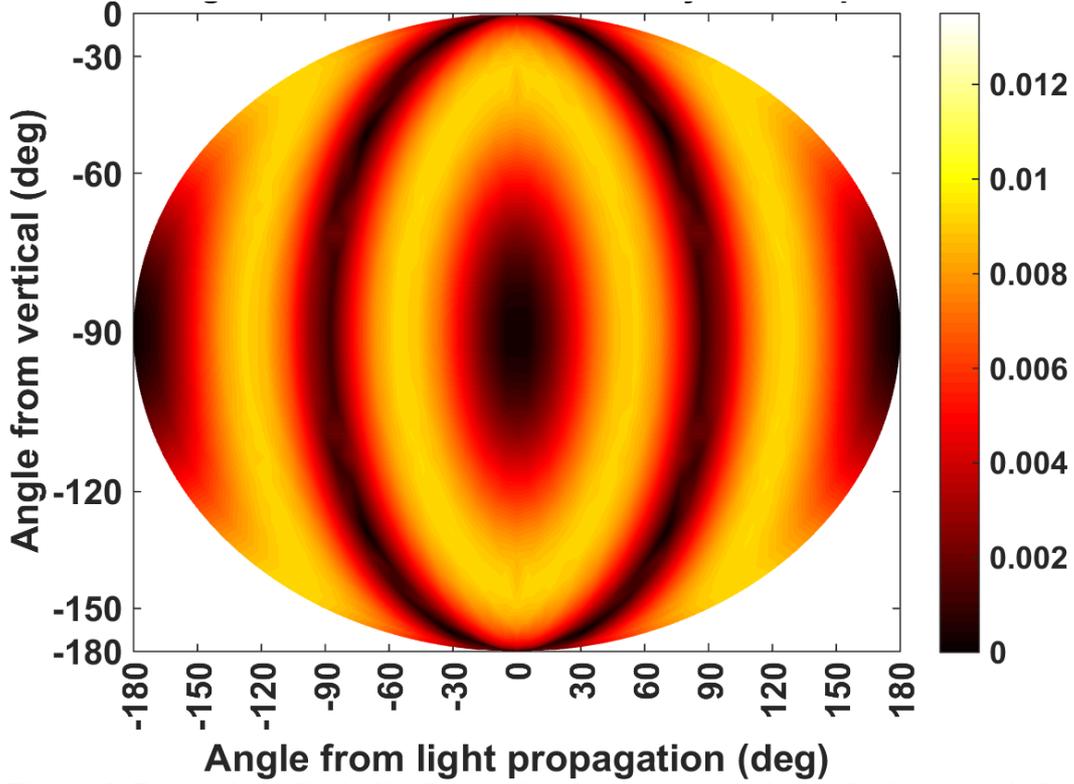

**Figure 10:** Expected amplitude distribution (in arbitrary units), calculated using numerical integration of Eq. 1.

## VIII. Conclusions

We have described the hardware and calibration techniques developed for experimental control of $B_0$ magnitude, orientation and homogeneity in a double-resonance magnetometer system. Through iterative optimisation of null-field compensation and coil calibration, we demonstrate tolerances of $\delta |B| = 0.94$ nT in $B_0$ magnitude and $\delta \theta = 5.9$ mrad in orientation for $B_0 = 200$ nT. By measuring magnetic resonance width, we can iteratively minimise inhomogeneity effects in this field, resolving the minimum with $\delta |\nabla B| = 13.0$ pT/mm.

This evaluation of precision and accuracy in $B_0$ allows further investigations to be conducted with known tolerances in field definition. Of particular interest to the authors is anisotropy in the response of double-resonance-type magnetometers with **$B_0$** orientation. It can be seen from Figures 9 and 10 that, even in the simple case of $M_x$ magnetometry with weak optical pumping, the observed anisotropic response is not easily reproduced by simulation.

Alkali vapour cell magnetometers are reaching technological maturity, and the benefits of their enhanced sensitivity will be reaped in a wide range of proposed applications. The development of detailed understanding of the effects of $B_0$ orientation on their sensitivity and signal response can be furthered by empirical testing utilising the precise field control demonstrated in this work.

**Acknowledgements**

The authors would like to thank Prof. Antoine Weis and Dr. Victor Lebedev of Fribourg University for supplying the Cs vapour cell used in this work. This work was funded by the UK Quantum Technology Hub in Sensing and Metrology, EPSRC (EP/M013294/1).